\documentclass[doublecol]{epl2} 
\title{Stokes drift for inertial particles transported by water waves}
\shorttitle{Stokes drift for inertial particles transported by water waves}

\author{G. Boffetta\inst{1} \and M. Martins Afonso\inst{2} \and 
A. Mazzino\inst{3,4} \and M. Onorato\inst{1} \and F. Santamaria\inst{1}}
\shortauthor{G. Boffetta, M. Martins Afonso, A. Mazzino, M. Onorato,
 F. Santamaria}

\institute{                    
\inst{1} Dipartimento di Fisica and INFN, Universit\`a di Torino, 
via Pietro Giuria 1, 10125 Torino, Italy \\
\inst{2} Institut de Math\'ematiques et de Mod\'elisation de Montpellier,
CNRS UMR 5149, Universit\'e Montpellier 2, c.c.051,
34095 Montpellier cedex 5, France \\
\inst{3} DICCA, Universit\`a di Genova, via Montallegro 1, 16145 Genova, Italy \\
\inst{4} INFN and CINFAI, Sezione di Genova, via Dodecaneso 33, 
16146 Genova, Italy
}
\pacs{47.35.-i}{Hydrodynamic waves}
\pacs{47.51.+a}{Mixing}
\abstract{
We study the effect of surface gravity waves on the motion of inertial
particles in an incompressible fluid.
We perform analytical calculations based on perturbation expansion which
allows us to predict the dynamics of inertial particles in deep water 
regime.
We find that the presence of inertia leads to a non-negligible correction to
the well-known horizontal Stokes drift velocity.
Moreover, we find that the vertical sedimentation velocity is also affected 
by a drift induced by waves. The latter result may have some relevant
consequences on the rate of sedimentation of particles of finite size. We
underline that the vertical drift would also be observed in the (hypothetical)
absence of the gravitational force.
Kinematic numerical simulations are performed and the results 
are found to be in excellent agreement with the analytical predictions, 
even for values of parameters beyond the perturbative limit.
}

\begin{document}

\maketitle

%%%%%%%%%%%%%%%%%%%%%%%%%%%%%%%%%%%%%%%%%%%%%%%%%%%%%%%%%%%%%%%%%%%%%%

\section{Introduction}
\label{sec:0}
The study of the Stokes drift velocity is a problem of paramount importance
both from a fundamental point of view \cite{V99} and in connection with 
applications, especially in the area of sediment transport 
\cite{L53,N92,VB96,BBV02,BVBLP12}.
As far as the first point is concerned, the Stokes drift velocity 
is for instance responsible of important fluid-mixing processes, 
including the mass and momentum transport near the free surface 
and the vertical-mixing enhancement owing to turbulent kinetic-energy 
production \cite{KWSC09}.
In the ocean, the Stokes drift is thought to be one important 
factor responsible for the Langmuir circulation \cite{MSM97}.
In relation to applications, it is known that
an accurate evaluation of the Stokes drift velocity is important
for the correct representation of surface physics in ocean general
circulation models and ocean models at smaller scales.
Other relevant effects on the ocean circulation are discussed, e.g., by 
\cite{N92}.

Since the seminal paper by \cite{S47}, Stokes drift has been recognized
as an important example that illustrates the difference between the 
Eulerian and the Lagrangian statistics \cite{L86}. 
It predicts that a fluid particle (i.e. a tracer
of negligible inertia) experiences a mean drift in the direction of
wave propagation proportional to $U^2/c$, where $U$ is the amplitude
of the wave-induced velocity and $c$ is the wave phase velocity.
Because the Stokes drift arises from the average of the wave motion
along a Lagrangian trajectory, 
it is relevant for all floating and suspended particles present in the
water column, and not only for fluid particles considered in the 
original derivation.
Inertia of finite-size particles with density different from the fluid
modifies Lagrangian trajectories with respect those of fluid particles.
This has important consequences on particle dispersion in both laminar 
and turbulent flows
\cite{SE91,WMBS92,FP04,BDG04,BBBCCLMT06,MA08,MAMO09,BBLST10,MAMM12,MM11},
and therefore we expect that inertia
might affect the Stokes drift experienced by inertial particles.
Previous studies in the field have investigated the case of particles
close to be neutrally buoyant in a velocity field generated by
internal gravity waves \cite{GRKZZ00} and small particles 
in deep water in the presence of surface gravity waves \cite{E08}.

Our main aim here is to push forward the analyses performed by these previous
studies and to investigate the role of inertia on the Stokes drift
velocity for particles of arbitrary density in deep-water waves.
As a result of our analysis, we show that inertia induces a correction to the
horizontal Stokes drift velocity which is a second-order effect 
in particle inertia, and generates a vertical drift (at the first
order in inertia) which modifies the sedimentation velocity in still
fluid.
We show that this vertical drift has a dynamical origin
as it is active even in the absence of gravity, a remarkable result 
not pointed out in previous studies.
The analytical results carried out by means of perturbative expansions
are corroborated by a set of numerical simulations
which extend the range of validity of our results beyond
the perturbative regime.

%The remaining of this paper is organized as follow. In Section~\ref{sec:2}
%we report the analytical calculations and results based on the 
%perturbation expansion. 
%Section~\ref{sec:3} is devoted to the numerical results and their
%comparison with analytical predictions, while 
%Section~\ref{sec:4} is devoted to conclusions.

%%%%%%%%%%%%%%%%%%%%%%%%%%%%%%%%%%%%%%%%%%%%%%%%%%%%%%%%%%%%%%%%%%%%%%
\section{Analytical study of inertial-particle motion}
\label{sec:2}

The Stokes drift velocity is a second-order effect in the
wave amplitude. Therefore, in order to be consistent in the
perturbative expansion, one has to consider
at least a second-order expansion for the wave motion, i.e.
a Stokes wave. In arbitrary water depth the Stokes wave 
generates the following two-dimensional, irrotational and 
incompressible velocity field $\bm{u}=(u,w)$ \cite{W74,KC90}
\begin{eqnarray}
u&=&U{\cosh[k(z+h)] \over \sinh(kh)}\cos(kx-\omega t)+ \nonumber \\
&& {3 U^2 \over 4 c}{\cosh[2k(z+h)] \over \sinh^4(kh)} 
\cos[2(kx-\omega t)] \label{eq:2.1} \\
w&=&U{\sinh[k(z+h)] \over \sinh(kh)}\sin(kx-\omega t)+ \nonumber \\
&& {3 U^2 \over 4 c}{\sinh[2k(z+h)] \over \sinh^4(kh)} 
\sin[2(kx-\omega t)]\;,
\label{eq:2.2}
\end{eqnarray}
where $x$ and $z$ are the horizontal and the vertical coordinate,
$h$ is the water depth, 
$k$ is the wave number and $\omega$ is the angular frequency
related to $k$ via the dispersion relation, $\omega=\sqrt{g k \tanh(k h)}$
(strictly speaking, one should include the nonlinear correction to the
dispersion relation; however, this turns out to be inessential in our analysis).
The phase velocity is $c=\omega/k$,
and the maximum velocity $U$ at the surface ($z=0$) of the first-order
solution is related to the wave amplitude $A$ by $U=\omega A$.
The equations (\ref{eq:2.1}) are obtained under the hypothesis
of small steepness $\epsilon=kA$, which, because of the relation between
$U$ and $A$, is equivalent to the Froude number defined as $Fr=U/c$.
Note that, in equations (\ref{eq:2.1}), the mean flow that should appear at the
same order as the second harmonic is not included because we are dealing with a
monochromatic wave, and not with wave packets characterized by modulation
length (see e.g. \cite{W74}, p.~474 for a discussion).

In the following we will consider the limit of deep water, $k h \to \infty$,
for which the coefficients of the second order terms in 
(\ref{eq:2.1}-\ref{eq:2.2}) vanish and the velocity field simplifies to
\begin{equation}
\bm{u}(x,z,t)=\left(U e^{k z}\cos(kx-\omega t),
U e^{kz}\sin(kx-\omega t)\right)\;.
\label{eq:2.6}
\end{equation}
We remark that the limit of deep water, which simplifies the following
analysis, is already valid with good approximation for $k h \simeq 2$
\cite{KC90}.
In this limit the dispersion relation reduces to $\omega=\sqrt{g k}$.

The motion of a small inertial particle transported by the fluid flow $\bm{u}$
through the Stokes drag (with typical response time $\tau$) and subject
to gravity acceleration $\bm{g}$ is given by
\begin{eqnarray}
{d \bm{x} \over d t} &=& \bm{V} 
\label{eq:2.3} \\
{d \bm{V} \over d t} &=& {\bm{u}-\bm{V} \over \tau} + (1-\beta) \bm{g}
+ \beta {d \bm{u} \over d t}\;, 
\label{eq:2.4}
\end{eqnarray}
where $\bm{x}(t)$ and $\bm{V}(t)$ represent the particle position and
velocity. In (\ref{eq:2.4}) the added-mass effect has been
taken into account via the dimensionless number
$\beta=3 \rho_\mathrm{f}/(\rho_\mathrm{f} + 2 \rho_\mathrm{p})$,
built from the fluid, $\rho_\mathrm{f}$,
and particle, $\rho_\mathrm{p}$, densities \cite{MR83,G83}.

In order to have an explicit expression for the particle velocity,
we expand (\ref{eq:2.4}) perturbatively in $\tau$ \cite{BFF01}, to obtain:
\begin{equation}
\bm{V}=\bm{u} + \tau (1-\beta) \left(\bm{g}-{d \bm{u} \over d t} \right)+
\tau^2 (1-\beta){d^2 \bm{u} \over d t^2} + O(\tau^3)\;.
\label{eq:2.5}
\end{equation}

%Besides the steepness,
%another dimensionless number, $\Gamma \equiv (1-\beta) g \tau c/U^2$,
%can be introduced as the ratio of the bare gravitational
%sedimentation velocity, $(1-\beta) g \tau$, over the drift velocity, $U^2/c$.
%Note that this number can be written as
%$\Gamma =(1-\beta) \mathrm{St}/\mathrm{Fr}^2$,
%where we have introduced the Stokes number
%$\mathrm{St} \equiv \omega \tau$.
%Indeed, since the velocity field (\ref{eq:1.5}) decays exponentially
%with $z$, the effect of the Stokes drift is modulated by the sedimentation
%process. Therefore, in the perturbation expansion, one has to decide at
%which order in $\epsilon$ the gravitational term has to be taken
%into account.
%We have analyzed different, physically relevant, cases
%and we found that the most interesting regime is when
%$\Gamma=O(1)$.
%Physically, this choice corresponds to a situation in which gravity is
%relatively small and particles experience a finite horizontal
%Stokes drift during sedimentation.
%In other words, the particle settling velocity in still fluids,
%$(1-\beta) g \tau$, made dimensionless via the wave velocity $c$,
%is assumed to be asymptotically $\propto U^2/c^2$,
%i.e.\ $O(\epsilon^2)$.

By introducing the dimensionless variables
$\bm{x}\mapsto k\bm{x}$, $t\mapsto\omega t$ and $\bm{u}\mapsto\bm{u}/U$,
we can expand $\bm{u}$ and its Lagrangian derivatives at the second 
order in $\epsilon$ and, by substitution in (\ref{eq:2.3})--(\ref{eq:2.5}),
we obtain for the particle motion:
\begin{equation}
\dot{x}=\epsilon \left[u-\mathrm{St}\beta'w-\mathrm{St}^2\beta'u \right]+ 
\epsilon^2 \mathrm{St}^2 \beta' e^{2z} + ... 
\label{eq:2.7}
\end{equation}
\begin{equation}
\dot{z}=-\mathrm{St} \beta' + 
\epsilon \left[w+\mathrm{St}\beta'u-\mathrm{St}^2 \beta \beta'w \right]-
\epsilon^2 \mathrm{St}\beta' e^{2z} + ... 
\label{eq:2.8}
\end{equation}
where we have introduced $\beta'\equiv 1-\beta$ and the Stokes number
$\mathrm{St} \equiv \omega \tau$.
We observe that the term $-\mathrm{St} \beta'$ represents the sedimentation
velocity in still fluid.

By expanding perturbatively the coordinate as
$\bm{x}=\bm{x}_0+\epsilon \bm{x}_1+\epsilon^2 \bm{x}_2 + ...$
and inserting into (\ref{eq:2.7}-\ref{eq:2.8})
we obtain a set of equations for the different orders in $\epsilon$. 
At order $\epsilon^0$ we simply have
\begin{equation}
\begin{array}{l}
x_0=x^* \\
z_0=z^*-\mathrm{St}\beta' t
\end{array}
\label{eq:2.9}
\end{equation}
where $(x^*,z^*)$ represents the initial position of the tracer.
At order $\epsilon$, after integration and taking
up to order $St^2$, we obtain
\begin{eqnarray}
x_1&=&-[1-\mathrm{St}^2\beta']e^{-\mathrm{St}\beta't}
\left[w^*+\mathrm{St}\beta'u^* \right]- \nonumber \\
&&\mathrm{St}\beta'e^{-\mathrm{St}\beta't}
\left[u^*-\mathrm{St}\beta'w^* \right]
\label{eq:2.10}
\end{eqnarray}
\begin{eqnarray}
z_1&=&-[1-\mathrm{St}^2\beta']e^{-\mathrm{St}\beta't}
\left[u^*-\mathrm{St}\beta'w^* \right]- \nonumber \\
&&\mathrm{St}\beta'e^{-\mathrm{St}\beta't}
\left[w^*+\mathrm{St}\beta'u^* \right]
\label{eq:2.11}
\end{eqnarray}
where ${\bf u}^*={\bf u}({\bf x}^*,t)$.
As in the original derivation for fluid particles, 
no drift velocity appears at the first order, and we need to 
go to the next order $\epsilon^2$. By substituting 
(\ref{eq:2.9}-\ref{eq:2.11}) into (\ref{eq:2.7}-\ref{eq:2.8}) 
and finally going back to original dimensional variables we obtain
\begin{eqnarray}
{dx \over dt}&=&U e^{k z_0(t)} \left[(1-\mathrm{St}^2 \beta' \beta)
\cos \phi^*(t) - \mathrm{St} \beta' \sin \phi^*(t) \right] + \nonumber \\
&& {U^2 \over c} e^{2 k z_0(t)} 
\left[1-\mathrm{St}^2 \beta' \beta \right]
\label{eq:2.12}
\end{eqnarray}
\begin{eqnarray}
{dz \over dt}&=&U e^{k z_0(t)} \left[(1-\mathrm{St}^2 \beta' \beta)
\sin \phi^*(t) + \mathrm{St} \beta' \cos \phi^*(t) \right] - \nonumber \\
&& \mathrm{St} \beta' \left[1+2 {U^2 \over c^2} e^{2 k z_0(t)} \right]
\label{eq:2.13}
\end{eqnarray}
where $\phi^*(t)=k x^*-\omega t$ is the Eulerian phase at the initial 
position and $z_0(t)=z^*-\mathrm{St}\beta' t$.
Now we assume that the bare sedimentation velocity $\mathrm{St}\beta'$ is small 
with respect to the wave motion, which is consistent with the 
expansion leading to (\ref{eq:2.5}), so that 
we can take $e^{k z_0}$ constant over a wave period.
Under this assumption, the 
drift velocities are simply given by the last terms in 
(\ref{eq:2.12}-\ref{eq:2.13}) which do not vanish when averaged 
over one period
\begin{equation}
u_\mathrm{d} = {U^2 \over c} \left[1-\beta(1-\beta) \mathrm{St}^2 \right]
e^{2k[z^*-(1-\beta)g \tau t]}\;,
\label{eq:2.14}
\end{equation}
\begin{equation}
w_\mathrm{d} = -(1-\beta) g \tau - 2 (1-\beta) \mathrm{St} {U^2 \over c}
e^{2k[z^*-(1-\beta)g \tau t]}\;.
\label{eq:2.15}
\end{equation}
In the limit of tracers, $\mathrm{St}=0$ and/or neutrally-buoyant particles
$\beta=1$, the above
expressions recover the velocities derived by Stokes:
$u_\mathrm{d}=e^{2 k z_0}U^2/c$, $w_\mathrm{d}=0$.
Inertia induces a correction of order $\mathrm{St}^2$ to this horizontal
drift velocity. The interesting result is that
inertia produces a drift velocity also in the vertical direction,
which corrects the bare sedimentation velocity. The correction
has the same sign of the velocity induced by gravity, i.e. negative (positive)
for heavy (light) particles. 
To our knowledge, the correction to sedimentation velocity induced
by water waves on inertial particles has never been discussed before.

It is interesting to observe that wave motion induces a vertical velocity 
also in the, unrealistic, case of $g=0$. 
Of course, the limit $g=0$ must be taken while keeping $k$ and $\omega$
independent (i.e. not related by the dispersion relation), otherwise 
$g=0$ would trivially imply $\omega=0$ and therefore no wave motion.
This vertical mean motion is produced by the combining effect of vertical 
symmetry breaking due to the z-dependence of the velocity field and 
the delayed dynamics induced by inertia.
By repeating the calculation with $g=0$ we obtain
\begin{equation}
u_\mathrm{d} = {U^2 \over c} \left[1-(2-\beta)(1-\beta) \mathrm{St}^2 \right]
e^{2kz^*}\;,
\label{eq:2.16}
\end{equation}
\begin{equation}
w_\mathrm{d} = - (1-\beta) \mathrm{St} {U^2 \over c}
e^{2kz^*}\;.
\label{eq:2.17}
\end{equation}

A simple, and physically relevant, prediction one can derive from
(\ref{eq:2.14}-\ref{eq:2.15}) is the net
displacement of particles from the point of release. For heavy
particles ($\beta<1$) released at the surface ($z_0=0$), a time integration
of (\ref{eq:2.14}) from $0$ to $\infty$ gives for the total displacement:
\begin{equation}
\Delta x=
%{U^2 \over 2 k (1-\beta) g \tau c} \left[1 - \beta (1-\beta) \mathrm{St}^2\right]=
{\mathrm{Fr}^2 \over 2 k} {1 - \beta (1-\beta) \mathrm{St}^2 \over (1-\beta) \mathrm{St}}\;.
\label{eq:2.18}
\end{equation}
Of course this expression can be valid only for
small $\mathrm{St}$, for which we have $\Delta x>0$.

%%%%%%%%%%%%%%%%%%%%%%%%%%%%%%%%%%%%%%%%%%%%%%%%%%%%%%%%%%%%%%%%%%%%%%
\section{Numerical simulations}
\label{sec:3}

In this Section we report the numerical results obtained from the integration
of (\ref{eq:2.3}) and (\ref{eq:2.4}), with the aim of verifying the analytical
predictions of the previous Section, and also to check the robustness of
these predictions with respect to the expansion parameters.

%%%%%%%%%%%%%%%%%%%%%%%%%%%%%%%%%%%%%%%%%%%%%%%%%%%%%%%%%%%%%%%%%%%%%%
\begin{figure}
\includegraphics[width=8.5cm]{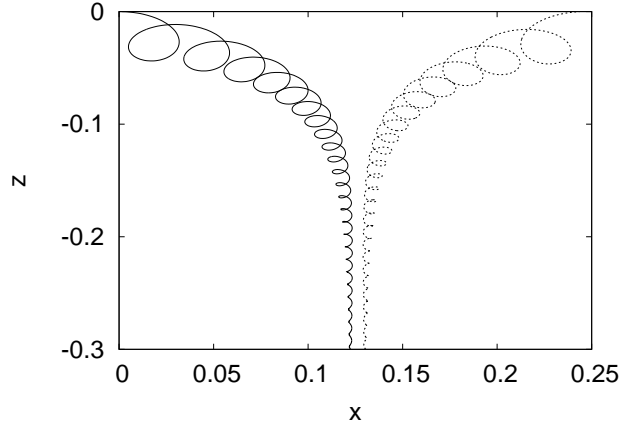}
\caption{Two examples of trajectories of slightly heavy ($\beta=0.99$,
continuous line) and light ($\beta=1.01$, dotted line) particles
obtained by numerical integration of (\ref{eq:2.3}-\ref{eq:2.4}) 
in the velocity field (\ref{eq:2.6}) with
$\epsilon=\mathrm{Fr}=0.33$ and $\mathrm{St}=0.5$.
The initial position for particles are $x^*=0$, $z^*=0$ (heavy)
and $x^*=0.13$, $z^*=-0.3$ (light).}
\label{fig1}
\end{figure}
%%%%%%%%%%%%%%%%%%%%%%%%%%%%%%%%%%%%%%%%%%%%%%%%%%%%%%%%%%%%%%%%%%%%%%

%%%%%%%%%%%%%%%%%%%%%%%%%%%%%%%%%%%%%%%%%%%%%%%%%%%%%%%%%%%%%%%%%%%%%%
\begin{figure}
\includegraphics[width=8.5cm]{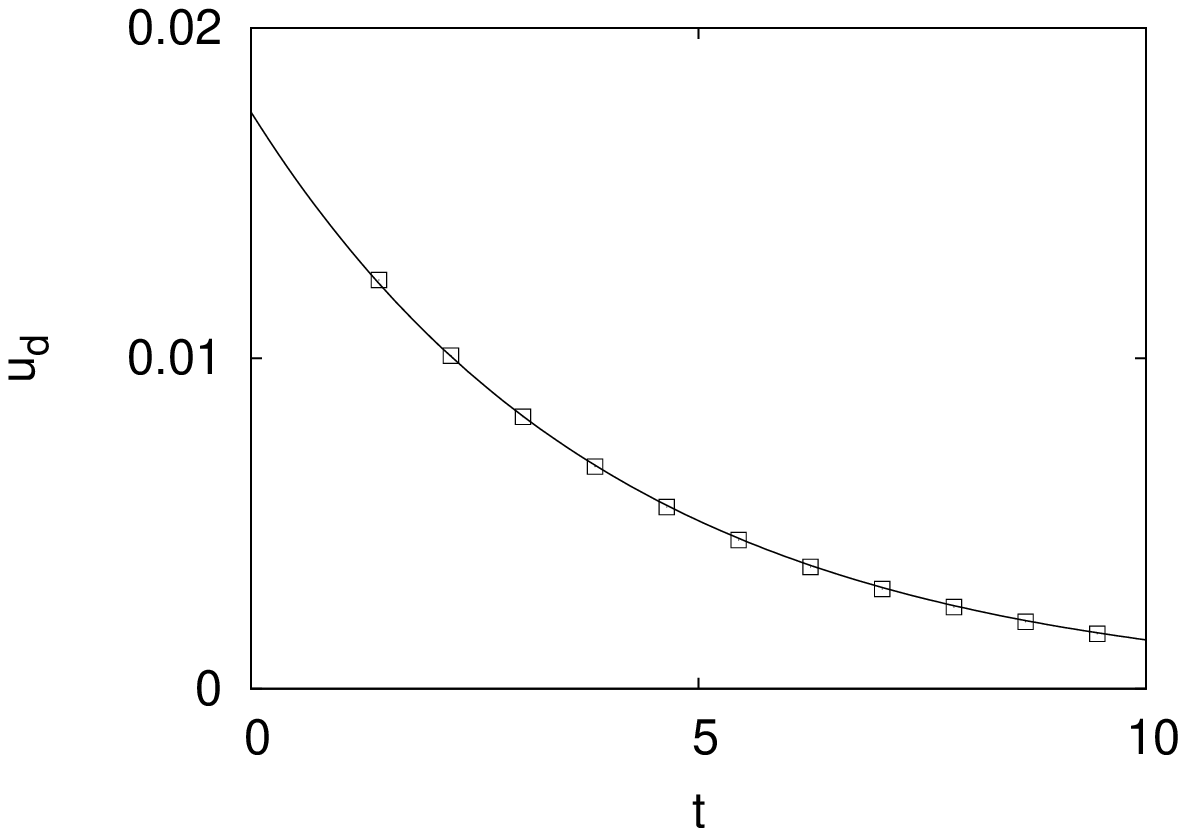}
\includegraphics[width=8.5cm]{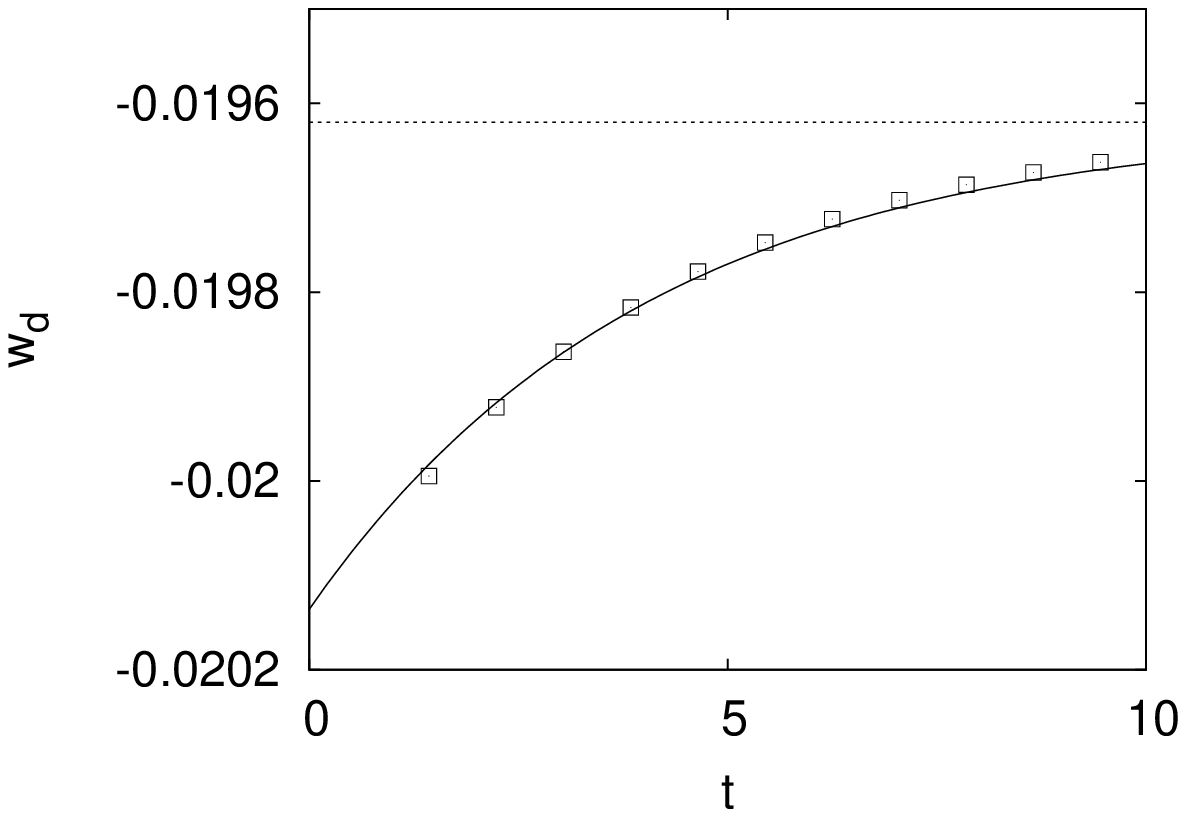}
\caption{Numerical (squares) vs.\ theoretical (solid line, equations
(\ref{eq:2.14}-\ref{eq:2.15})) drift velocity:
horizontal (upper panel) and vertical (lower panel) components. The dashed
line represents the settling velocity in the absence of wave motion
$w_d=-(1-\beta) g \tau$.
Parameters: $\epsilon=\mathrm{Fr}=0.125$, $\mathrm{St}=0.157$, $\beta=0.9$.}
\label{fig2}
\end{figure}
%%%%%%%%%%%%%%%%%%%%%%%%%%%%%%%%%%%%%%%%%%%%%%%%%%%%%%%%%%%%%%%%%%%%%%

%%%%%%%%%%%%%%%%%%%%%%%%%%%%%%%%%%%%%%%%%%%%%%%%%%%%%%%%%%%%%%%%%%%%%%
\begin{figure}
\includegraphics[width=8.5cm]{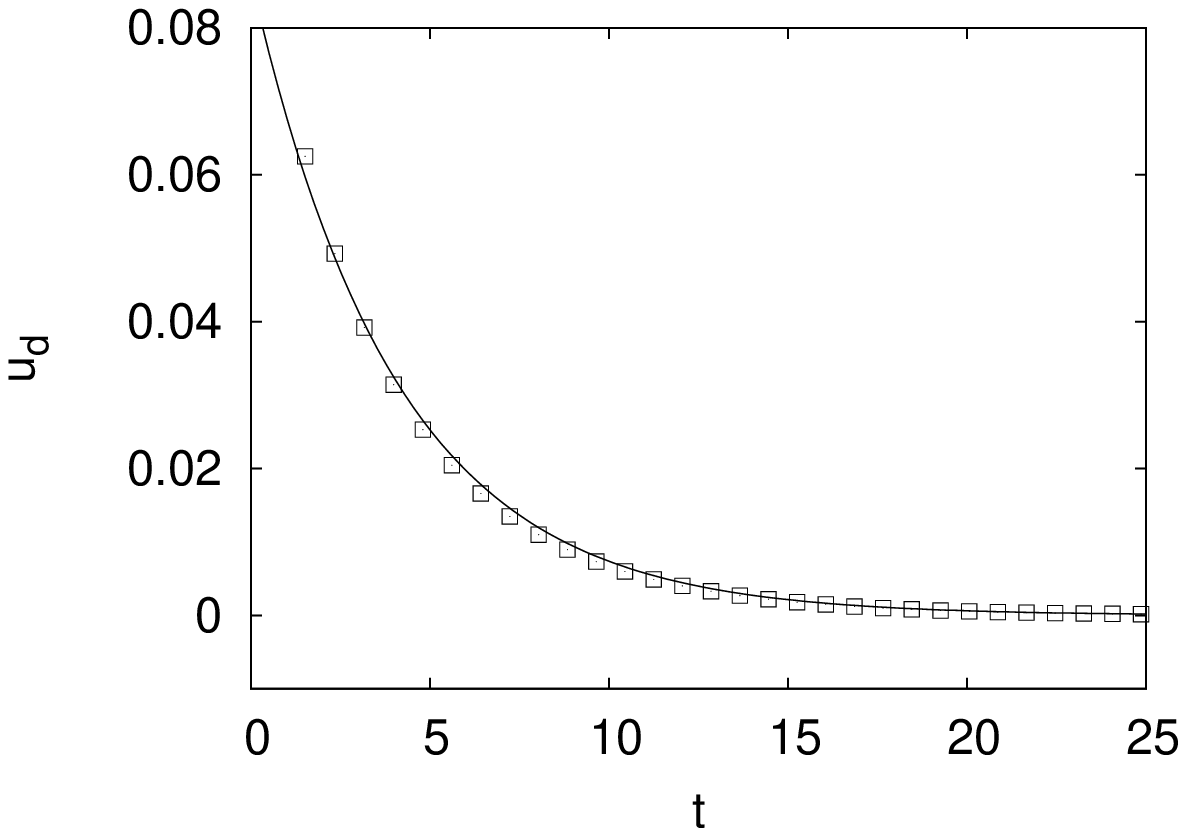}
\includegraphics[width=8.5cm]{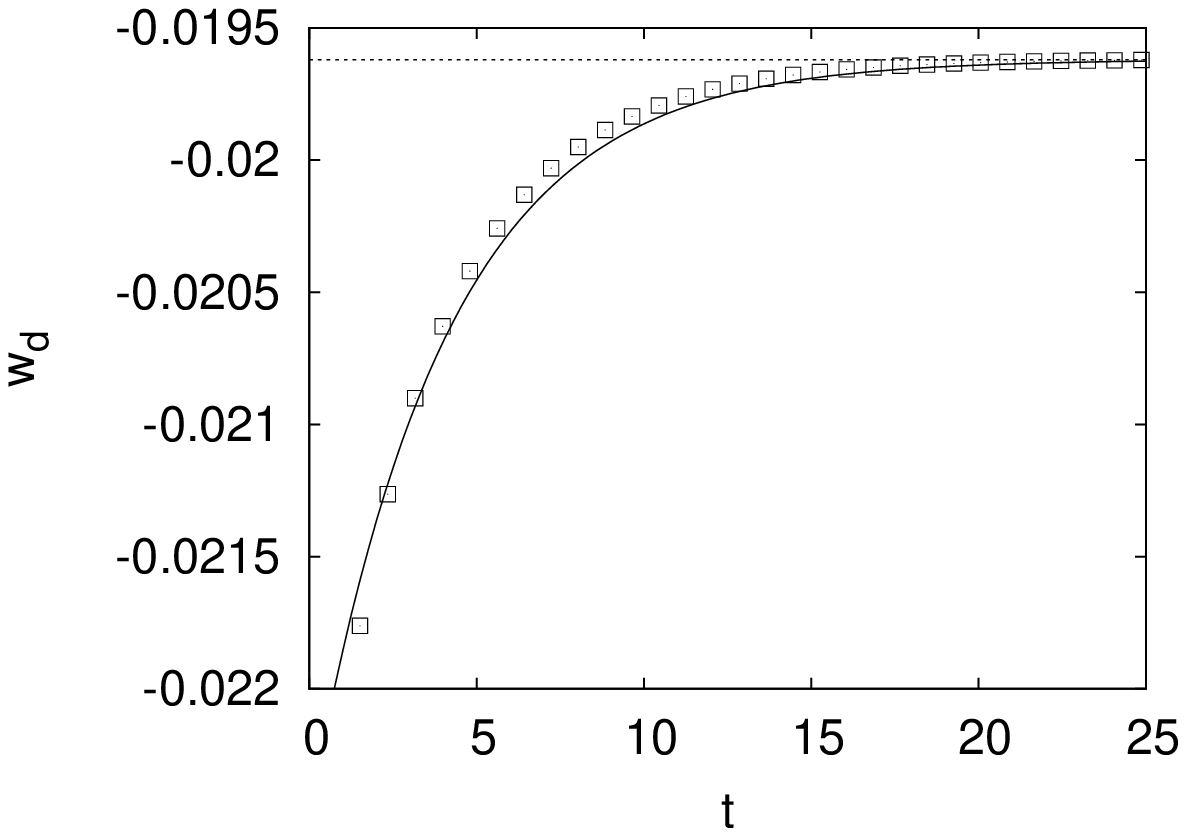}
\caption{Numerical (squares) vs.\ theoretical (solid line) drift velocity:
horizontal (upper panel) and vertical (lower panel) components. The dashed
line represents the settling velocity in the absence of wave motion
$w_d=-(1-\beta) g \tau$.
Parameters: $\epsilon=\mathrm{Fr}=0.33$, $\mathrm{St}=0.157$, $\beta=0.9$.}
\label{fig3}
\end{figure}
%%%%%%%%%%%%%%%%%%%%%%%%%%%%%%%%%%%%%%%%%%%%%%%%%%%%%%%%%%%%%%%%%%%%%%

Figure~\ref{fig1} shows two typical examples of trajectories of slightly heavy
and light particles induced by linear waves in deep water. 
From these trajectories the drift velocity is obtained by computing 
the horizontal and vertical displacement of the position over one
Lagrangian period (defined from the point at which the Lagrangian 
horizontal velocity changes from negative to positive) divided by
the period.
An example of the resulting velocity components is shown in Fig.~\ref{fig2}
for a wave with $\epsilon=0.125$. For this case at moderate steepness
the agreement with the theoretical prediction (\ref{eq:2.14}) and
(\ref{eq:2.15}) is very good. From the plot of the vertical velocity
$w_d$ we see that the relative correction induced by
waves to the bare sedimentation velocity at the initial time is 
$2 \mathrm{Fr}^2\approx 0.03$.

%%%%%%%%%%%%%%%%%%%%%%%%%%%%%%%%%%%%%%%%%%%%%%%%%%%%%%%%%%%%%%%%%%%%%%
\begin{figure}
\includegraphics[width=8.5cm]{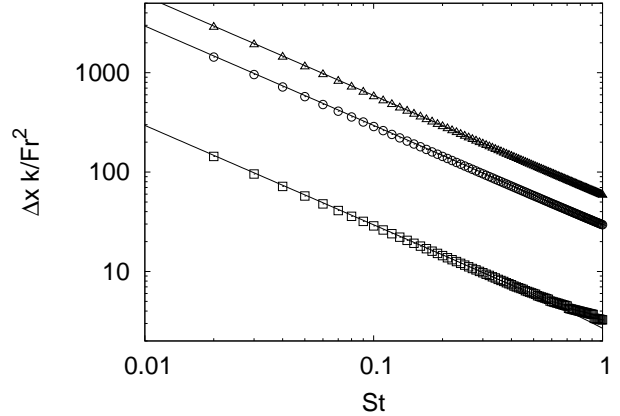}
\caption{Total horizontal displacement of heavy particles
with $\beta=0.9$ (squares), $\beta=0.99$ (circles) and $\beta=0.995$
(triangles), settling
beneath a linear wave with $\epsilon=\mathrm{Fr}=0.33$, as a function of $St$.
Lines represent theoretical predictions (\ref{eq:2.18}).}
\label{fig4}
\end{figure}
%%%%%%%%%%%%%%%%%%%%%%%%%%%%%%%%%%%%%%%%%%%%%%%%%%%%%%%%%%%%%%%%%%%%%%

For steeper waves, as the example shown in Fig.~\ref{fig3} at
$\epsilon=0.33$, the agreement between numerical simulations and
perturbative predictions worsens, nonetheless 
(\ref{eq:2.14}) and (\ref{eq:2.15}) still give a good approximation 
of the numerical data, with a correction to the
bare sedimentation velocity at initial time of about $0.2$. 
Of course for even larger steepness other effects such
as wave breaking, clearly not included in the theory or simulation,
can take place.

The total horizontal displacement of heavy particles released at the
surface of deep water for different values of the parameters is shown
in Fig.~\ref{fig4}, together with the predictions given by
(\ref{eq:2.18}). The agreement is very good not only in the perturbative
regime of small $\mathrm{St}$ in which (\ref{eq:2.18}) is derived.
Deviations are observable only for $\mathrm{St}=O(1)$.

%%%%%%%%%%%%%%%%%%%%%%%%%%%%%%%%%%%%%%%%%%%%%%%%%%%%%%%%%%%%%%%%%%%%%%
\begin{figure}
\includegraphics[width=8.5cm]{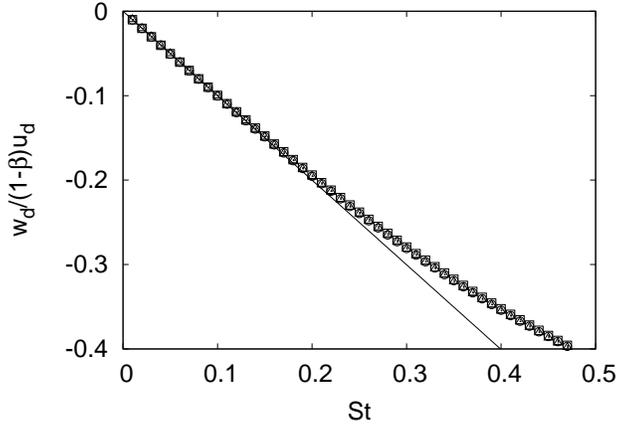}
\caption{Ratio of vertical to horizontal drift velocity
$w_d/(1-\beta)u_d$ as a function of $St$ for heavy particles
in the absence of gravity ($g=0$). Parameters: $\epsilon=0.33$,
$\beta=0.9$ (squares), $\beta=0.99$ (circles), $\beta=0.995$ (triangles).
The line represents the prediction
$w_d/(1-\beta)u_d=-St/[1-(2-\beta)(1-\beta)St^2]$.}
\label{fig5}
\end{figure}
%%%%%%%%%%%%%%%%%%%%%%%%%%%%%%%%%%%%%%%%%%%%%%%%%%%%%%%%%%%%%%%%%%%%%%

As discussed in the previous Section, the perturbative analysis shows 
that inertial particles have a vertical drift even in the absence
of gravity. This drift acts in the same direction of sedimentation
as it has the same sign of the gravitational term, independently
of $\beta$. By comparing (\ref{eq:2.16}) and (\ref{eq:2.17}), we see 
that for $g=0$ the mean motion is along a straight line, with slope given by
$w_\mathrm{d}/u_\mathrm{d}=
-(1-\beta)\mathrm{St}/[1-(2-\beta)(1-\beta)\mathrm{St}^2]$. 
We observe that in this limit the total displacement $\Delta x$
of a heavy particle in deep water diverges.
Figure~\ref{fig5} shows the slope of the mean motion as a function of
$\mathrm{St}$ for different values of $\beta$. Again, for small and
moderate values of the parameter, the agreement with the analytical
prediction is very good.

\section{Conclusions}
\label{sec:4}
We have considered the problem of Stokes drift induced by water waves
in deep water limit 
on small inertial particles with two complementary perspectives.
On the one hand, our results give the correction
to the horizontal Stokes drift induced
by inertia. This correction is found to be second order in the
particle Stokes number, with a sign which depends on the particle
density relative to water.
On the other hand, we also obtain a vertical drift velocity, which therefore
represents a correction to the sedimentation velocity induced by wave
motion on the surface. This effect, which results to be at the first order
in the Stokes number, has never been discussed before and is of possible
relevance, e.g., in the field of sediment transport in coastal regions.

We conclude by observing that the present analysis is performed
in the ideal world of linear two-dimensional water waves and
in the absence of any interactions with physical boundaries.
One can speculate whether our main findings will survive in
more complex and realistic situations. Because the results are
based on a simple kinematic model, we conjecture that 
the corrections induced by finite Stokes number will survive 
to more complex velocity field, at least at a qualitative level.
It would be therefore extremely interesting
to study the drift velocity of inertial particles, and its effect 
on sedimentation, in more realistic simulations of wave motion and in 
laboratory experiments, where a precise determination of the mean 
velocity and falling velocity is possible.

\acknowledgements
We thank D. Pugliese who participated in the preliminary phase of this
study.
The authors acknowledge the support from the EU COST Action MP0806 
``Particles in Turbulence''.

\bibliography{biblio}
\bibliographystyle{eplbib}

\end{document}